\documentclass[aps,prl,twocolumn,showpacs,superscriptaddress,letterpaper]{revtex4-1}
\usepackage{mathrsfs}
\usepackage{amsmath}
\usepackage{bm}
\usepackage{color}
\usepackage{lineno}
\usepackage{graphicx}
\usepackage{subfig}
\usepackage{hyperref}

\begin{document}
\title{How Zonal Flow Affects Trapped Electron Driven Turbulence in Tokamaks}

\author{Haotian Chen}
\affiliation{Institute of Space Science and Technology, Nanchang University, Nanchang, 330031, China}
\affiliation{Department of Atomic, Molecular and Nuclear Physics, University of Seville, Seville, 41012, Spain}
\author{Liu Chen}
\email[]{liuchen@zju.edu.cn}
\affiliation{Institute for Fusion Theory and Simulation and Department of Physics, Zhejiang University, Hangzhou, 310027, China}
\affiliation{Department of Physics and Astronomy, University of California, Irvine, California 92697, USA}
\affiliation{Center for Nonlinear Plasma Science and C.R. ENEA Frascati - C.P. 65, 00044 Frascati,Italy}

\vskip 0.25cm

\date{\today}

\begin{abstract}

	The role of self-generated zonal flows in the collisionless trapped-electron mode (CTEM) turbulence is a long-standing open issue in tokamak plasmas.
	Here we show, for the first time, that the zonal flow excitation in the CTEM turbulence is formally isomorphic to that in the ion temperature gradient turbulence. 
	Trapped electrons contribute implicitly only via linear dynamics. 
	Theoretical analyses further suggest that,  for short wavelength CTEMs, the zonal flow excitation is weak and, more importantly, not an effective saturation mechanism. Corresponding controlling parameters are also identified theoretically.
	These findings not only offer a plausible explanation for previous seemingly contradictory simulation results, but can also  facilitate controlling the CTEM instability and transport with experimentally accessible parameters.

\end{abstract}


\maketitle


	Zonal flows (ZFs) are azimuthally symmetric sheared flows spontaneously excited by small scale fluctuations.
	They are common in both nature and laboratory, and 
	are crucial for the self-regulation of turbulence and transport.
	In tokamak plasmas, the drift wave-zonal flow paradigm, as one of  major achievements of modern plasma turbulence theory, has been established after intensive studies of the ion temperature gradient (ITG) turbulence (see \cite{diamond05} and references therein).
	At present it has been widely accepted that ZF-induced scatterings from unstable to stable domain (in wavenumber space) of either dominant \cite{chen00} or subdominant \cite{makwana} branches provide the primary saturation mechanism for ITG turbulence.
	However, noting that fusion power increases as the square of density and fusion produced alpha particles mainly heat electrons, the collisionless trapped-electron mode (CTEM) turbulence may show more pronounced effects in future burning plasmas such as ITER \cite{shimada}.
	That is, in addition to playing a key role in particle transport, CTEM also significantly contributes to electron heat transport in ITER scenarios with dominant electron heating and/or with comparable electron and ion heating.
	Although that the CTEM turbulence is of theoretical and practical interests has long been recognized, the significance of ZF in regulating the CTEM turbulence, nevertheless, is still being actively debated.
Numerical studies have shown that the role of ZF in CTEM turbulence is parameter-sensitive, with different controlling parameters identified from different simulations \cite{ernst04, dannert,lang07,lang08,waltz08,ernst,jolliet,xiao09,xiao10, nakata}. 
	Specifically, the ZF excitation is empirically found to be sensitive either to  the ratio between electron and ion temperatures ($\tau$), magnetic shear ($s$) and  electron temperature gradient scale length ($r_{te}$) \cite{dannert,lang07,lang08,waltz08}; or to $\eta_{e}$ (the ratio between gradients of the density and electron temperature) only \cite{ernst}.
It is conjectured that the importance of ZF  may be connected with the linear stability of CTEM \cite{ernst}, but the underlying physics mechanism is as yet unknown. 
The need for a clear physical picture of CTEM-ZF interplay is thus the main motivation for this work.

	In this Letter, we employ the nonlinear gyrokinetic theory \cite{frieman} and  demonstrate analytically that the ZF excitation in CTEM turbulence is formally isomorphic to that in ITG turbulence. 
Interestingly, although the turbulence is driven by trapped electrons, the nonlinear CTEM-ZF interplay is governed by ions and circulating electrons. Trapped electrons, in contrast, only enter implicitly through linear physics.
	Therefore, linear CTEM properties play a unique role in determining the importance of ZF.
	Theoretical analysis elucidates that ZFs are important in saturating the long wavelength CTEM turbulence,
	consistent with simulation results \cite{ernst04,lang07,lang08,waltz08,ernst, nakata,xiao09,xiao10}.
	For the short wavelength CTEM turbulence (to be defined later), however, ZF excitation is weak and, more significantly, not an effective saturation channel.
Linear short wavelength CTEMs are thus revisited analytically.
It is found that the short wavelength CTEM instability without ZF scattering channel is essentially of two types. One is kinetically excited via toroidal precessional resonance. 
	In this case, the instability threshold depends on the aspect ratio between major and minor radii, the temperature ratio, magnetic shear and electron temperature gradient; consistent with previous numerical simulation observations in Refs. \cite{dannert,lang07,lang08,waltz08}.
The other case is a fluid-like interchange-driven instability set by $\eta_{e}$ in the steep density gradient regime, explaining thereby the simulation results in Ref. \cite{ernst}.
	We note that while this work analyzes CTEMs in tokamaks, the theoretical approach presented here could conceivably be applicable toward clarifying the role of ZF in other plasma turbulence; e.g., shear Alfv\'{e}n waves excited by either energetic \cite{chen16} or thermal \cite{heidbrink} particles,  where wave-particle interactions of different particle species  occur on distinctively separated spatial scales.

	We consider an axisymmetric, low-$\beta$ (the ratio of kinetic to magnetic pressure), large aspect-ratio ($\epsilon=r/R_{0}\ll 1$) tokamak with the major radius $R_{0}$, and the radial ($r$, distance from the magnetic axis), poloidal ($\theta$) and toroidal ($\zeta$) field-aligned coordinates.
	The electrostatic fluctuation is taken to be coherent and consists of the pump CTEM $A_{0}$, $(\omega_{0}, \bm{k}_{0})$,  upper and lower sideband CTEMs $A_{\pm}$, $(\omega_{\pm}, \bm{k}_{\pm}\equiv \bm{k}_{0}\pm k_{z}\hat{e}_{r})$, and a zonal mode $A_{z}$, $(\omega_{z}, k_{z}\hat{e}_{r})$.
For CTEMs, since the instability drive  peaks on the outboard midplane, 
we can assume eigenmodes are mainly formed within $|\theta|\le \pi$,
and adopt the strongly ballooning representation \cite{connor78}
\begin{eqnarray}
\label{eq:pump}
	\{\phi_{\bm{k}},\delta h_{\bm{k}}\}=A_{\bm{k}}e^{-inq\theta_{k}}e^{-in(\zeta-q\theta)}\{\Phi_{\bm{k}}(\theta),\delta H_{\bm{k}}(\theta)\}.
\end{eqnarray}
Here, $q(r)$ stands for the safety factor, $A_{\bm{k}}$ denotes the amplitude, and $nq'\theta_{k}$ is the radial envelope wavenumber. 
For simplicity, we take $\theta_{k}=0$ for the pump and $k_{z}=nq'\theta_{z}$ for the zonal mode, hence the radial envelope wavenumber of upper (lower) sideband becomes $\pm k_{z}$.
The nonlinear equation for the electrostatic potential $\phi_{\bm{k}}$ is the quasineutrality condition:
\begin{eqnarray}
\label{eq:quasineutrality}
	(1+\tau)\phi_{\bm{k}}+\langle\delta h_{e,\bm{k}}\rangle_{v}-\langle J_{\bm{k}}\delta h_{i,\bm{k}}\rangle_{v}=0,
\end{eqnarray}
where $\tau=T_{e}/T_{i}$ is the temperature ratio between electron and ion,  $\langle\cdots\rangle_{v}$ denotes integration in velocity space, and $J_{\bm{k}}=J_{0}(k_{\perp}\rho_{i}v_{\perp})$ is a Bessel function accounting for the finite Larmor radius (FLR) effect of ions, with $\rho_{i}$ being the thermal ion Larmor radius and velocity being normalized to the thermal velocity.
The nonadiabatic  response $\delta h_{j, \bm{k}}$, meanwhile,  obeys the nonlinear gyrokinetic equation \cite{frieman}
\begin{eqnarray}
\label{eq:nlgk}
	& &\mathcal{L}_{j,\bm{k}}\delta h_{j,\bm{k}}-(\omega+\omega_{*j}^{t}) F_{0}J_{\bm{k}}\frac{q_{j} \phi_{\bm{k}}}{T_{j}}\nonumber\\
	&=&\frac{ic}{B}\sum_{\bm{k}_{2}=\bm{k}-\bm{k}_{1}}\{[J_{\bm{k}_{1}} \phi_{\bm{k}_{1}},\delta h_{j,\bm{k}_{2}}^{*}]+[J_{\bm{k}_{2}} \phi_{\bm{k}_{2}}^{*},\delta h_{j,\bm{k}_{1}}]\}.
\end{eqnarray}
Here, $j=i,e$ denotes particle species, the charge $q_{i}=-q_{e}=e$,
 and $\mathcal{L}_{j,\bm{k}}=\omega+\omega_{t}+\omega_{dj}$ is the phase-space propagator in toroidal geometry, in which $\omega_{t}=i(v_{\parallel}/q R_{0})\partial_{\theta}$ is the transit frequency, and $\omega_{dj}=2k_{\theta}v_{\parallel}^{2}G/(\omega_{cj}R_{0})$ is the curvature drift model for magnetic drift frequency \cite{terry}, with $G\equiv \cos\theta+s(\theta-\theta_{k})\sin\theta$, $k_{\theta}=nq/r$ and $\omega_{cj}=q_{j}B/(c m_{j})$.
$\omega_{*j}^{t}=\omega_{*j}[1+\eta_{j}(v^{2}-3/2)]$ is the diamagnetic frequency, where $\omega_{*j}=k_{\theta}c T_{j}/(q_{j}B r_{n})$,  $\eta_{j}=r_{n}/r_{tj}$, and $r_{n}$ and $r_{tj}$ are, respectively, the density and temperature scale lengths. For clarity of the physics presentation,  we set $\eta_{i}=0$ to inhibit ion driven modes.
 $F_{0}$ is Maxwellian.
We have also defined the Poisson bracket $[f, g]=i(k_{\theta,f}f\partial_{r}g-k_{\theta,g}g\partial_{r}f)$ for the $\bm{E}\times\bm{B}$ nonlinearity on the right-hand side, with $k_{\theta,f}$ ($k_{\theta,g}$) denoting the poloidal wavenumber $k_{\theta}$ of $f$ ($g$).

For trapped electrons, Eq. (\ref{eq:nlgk}) can be further reduced to the nonlinear bounce kinetic equation \cite{gang}:
\begin{eqnarray}
\label{eq:bke}
	& &\mathcal{L}_{te,\bm{k}}\delta H_{te,\bm{k}}+(\omega+\omega_{*e}^{t})F_{0}\frac{e\overline{\Phi_{\bm{k}}}}{T_{e}}\nonumber\\
	&=&\frac{ic}{B}\sum_{\bm{k}=\bm{k}_{1}-\bm{k}_{2}}\{[\overline{\Phi_{\bm{k}_{1}}},\delta H_{te,\bm{k}_{2}}^{*}]+[\overline{\Phi_{\bm{k}_{2}}^{*}},\delta H_{te,\bm{k}_{1}}]\},
\end{eqnarray}
where 
$\overline{(\cdots)}=[\oint (\cdots) d\theta/v_{\parallel}]/\oint d\theta/v_{\parallel}$ denotes bounce/transit averaging \cite{tang}.
  The propagator $\mathcal{L}_{te,\bm{k}}=\omega-\bar{\omega}_{de}$,
with the precessional frequency $\overline{\omega_{de}}=|\omega_{*e}|\epsilon_{n}v^{2}H$, $\epsilon_{n}=r_{n}/R_{0}$ and $H\simeq 0.83s+0.41$ for $s\sim 1$  \cite{sm,adam} typically employed in simulations \cite{ernst04, dannert,lang07,lang08,waltz08,ernst,jolliet,xiao09,xiao10, nakata}.

Although linear CTEMs have been extensively explored in the literature since 1970s \cite{adam, chen18, coppi}, it is nevertheless  crucial to discuss the linear properties useful for nonlinear analysis.
For this, the eigenmode equation is derived from linearized kinetic equations  as \cite{chen19}:
\begin{eqnarray}
\label{eq:dispersionfunction}
	\hat{D}\Phi&\equiv&(1+\tau)\Phi-\langle J_{0}\mathcal{L}_{i}^{-1}[(\omega+\omega_{*i}^{t})\tau F_{0}J_{0}\Phi]\rangle_{v}\nonumber\\
	& &-\langle \mathcal{L}_{te}^{-1}[(\omega+\omega_{*e}^{t})F_{0}\overline{\Phi}]\rangle_{v}=0.
\end{eqnarray}
Equation (\ref{eq:dispersionfunction}) distinguishes two types of CTEMs. The long wavelength mode  ($k_{\perp}^{2}\rho_{i}^{2}\sim \mathcal{O}(\epsilon)$) comes from  a balance between the adiabatic term $(1+\tau)$  and the nonadiabatic ion response (the second term), which requires that the nonadiabatic trapped-electron response is subdominant; while the  short wavelength mode ($k_{\perp}^{2}\rho_{i}^{2}\sim \mathcal{O}(\epsilon^{-1})$) is defined here as the result of a balance between the adiabatic term and nonadiabatic trapped-electron response, implying a subdominant nonadiabatic ion contribution. 

As to the excitation of ZF by CTEM turbulence, we may follow the analyses of Refs. \cite{rosenbluth, chen00, zonca04}. 
The particle response to ZF can be solved from Eq. (\ref{eq:nlgk}), by employing the separation in time scales between the low frequency zonal flow and the bounce (or transit) frequency of guiding-center motion \cite{rosenbluth}.
The quasineutrality condition of zonal mode then becomes:
\begin{eqnarray}
\label{eq:nonlocal_Az}
(\partial_{t}+\gamma_{z})\chi_{i}A_{z}= k_{z}\rho_{i}(A_{+}A_{0}^{*}\beta_{z,+}-A_{0}A_{-}^{*}\beta_{z,-}^{*}).
\end{eqnarray}
Here, 
frequencies and amplitudes are normalized  to $|\omega_{*e}|$ and $(T_{e}/e)(\rho_{i}/r_{n})$, respectively.
$\gamma_{z}$ is the collisional ZF damping rate.
$\chi_{i}=\tau(1-\langle\langle  F_{0}J_{z}e^{iQ}\overline{e^{-iQ}}J_{z}\rangle\rangle_{v,s})$ quantifies the classical and neoclassical polarization of ions \cite{rosenbluth},  with $Q=k_{z}v_{\parallel}q/(\omega_{ci}\epsilon)$ accounting for the ion finite drift orbit width effect, and $\langle \cdots\rangle_{s}$ being the average over magnetic flux surface.
The nonlinear coupling coefficient is given by \cite{zonca04}
	$\beta_{z,\pm}=\langle\langle J_{k_{z}}e^{iQ}\overline{e^{-iQ} (J_{\bm{k}_{\pm}}\Phi_{\bm{k}_{\pm}}\delta H_{i,\bm{k}_{0}}^{*})}\rangle\rangle_{v,s}-\langle\langle \Phi_{\bm{k}_{\pm}}J_{\bm{k}_{0}}\delta H_{i,\bm{k}_{0}}^{*}\rangle\rangle_{v,s}-\langle\langle J_{k_{z}}e^{iQ}\overline{e^{-iQ} (J_{\bm{k}_{0}}\Phi_{\bm{k}_{0}}^{*}\delta H_{i,\bm{k}_{\pm}})}\rangle\rangle_{v,s}+\langle\langle  \Phi_{\bm{k}_{0}}^{*}J_{\bm{k}_{\pm}}\delta H_{i,\bm{k}_{\pm}}\rangle\rangle_{v,s}$.
Consistently with $\theta_{k}=0$ for the pump mode, one can show that $\beta_{z,+}=\beta_{z,-}\equiv\beta_{z}$ for sidebands of the same branch by virtue of the up-down symmetry \cite{chen18}: $\Phi_{\bm{k}_{+}}=\Phi_{\bm{k}_{-}}$ and $\omega_{\bm{k}_{+}}=\omega_{\bm{k}_{-}}$.
Since the characteristic CTEM frequency is much smaller than the electron transit frequency, the circulating electron response to CTEMs is adiabatic.
The corresponding electron contribution to  $\beta_{z,\pm}$ is thus linearly proportional to
$\langle\langle\overline{[\phi_{\bm{k}_{\pm}},\delta h_{te,\bm{k}_{0}}^{*}]}\rangle\rangle_{v,s}-\langle\langle [\phi_{\bm{k}_{\pm}},\delta h_{te,\bm{k}_{0}}^{*}]\rangle\rangle_{v,s}+\langle\langle\overline{[\phi_{\bm{k}_{0}}^{*},\delta h_{te,\bm{k}_{\pm}}]}\rangle\rangle_{v,s}-\langle \langle[\phi_{\bm{k}_{0}}^{*},\delta h_{te,\bm{k}_{\pm}}]\rangle\rangle_{v,s}$.
Meanwhile, recalling that the bounce kinetic limit holds for trapped electrons, it is possible to show that $\langle\langle\overline{[\phi_{\bm{k}_{1}},\delta h_{te,\bm{k}_{2}}^{*}]}\rangle\rangle_{v,s}=\langle\langle[\overline{\Phi_{\bm{k}_{1}}},\delta H_{te,\bm{k}_{2}}^{*}]\rangle\rangle_{v,s}=\langle\langle [\phi_{\bm{k}_{1}},\delta h_{te,\bm{k}_{2}}^{*}]\rangle\rangle_{v,s}$ \cite{gang}.
As a consequence, trapped electrons do not contribute to $\beta_{z,\pm}$.
The nonlinear terms recover the well-known Charney-Hasegawa-Mima nonlinearity  in long wavelength limit \cite{hasegawa}.
It is, therefore, clear that the ZF equation is only set by ion dynamics, as that in  the familiar ITG turbulence.
Specifically, assuming $k_{\perp0}^{2}\rho_{i}^{2}\sim k_{z}^{2}\rho_{i}^{2}\ll 1$, it is not hard to show that Eq. (\ref{eq:nonlocal_Az}) is mathematically identical to the ZF equation in \cite{chen00,zonca04}.

For CTEM sidebands, 
the nonlinear coupling to the pump and ZF can be straightforwardly derived  by closely following Refs. \cite{chen00, zonca04}, yielding
\begin{eqnarray}
\label{eq:Ap}
	(\partial_{t}+i\Delta_{+}-\gamma_{+})A_{+}=-k_{z}\rho_{i}\alpha_{+}A_{0}A_{z}/\omega_{+}\partial_{\omega}D_{+},
\end{eqnarray}
and
\begin{eqnarray}
\label{eq:An}
	(\partial_{t}+i\Delta_{-}-\gamma_{-})A_{-}=k_{z}\rho_{i}\alpha_{-}A_{0}A^{*}_{z}/\omega_{-}\partial_{\omega}D_{-}.
\end{eqnarray}
Here,
$\Delta_{\pm}=\omega_{\pm}-\omega_{0}$ is frequency mismatch, $\gamma_{\pm}$ is sideband linear damping rate,
and $D_{\bm{k}}$ represents the real part of linear dispersion function $\langle\Phi_{\bm{k}}^{*}\hat{D}\Phi_{\bm{k}}\rangle_{s}$.
The mode coupling $\alpha_{\pm}\equiv \alpha_{e\pm}+\tau\alpha_{i\pm}$ here originates from both the Navier-Stokes type $\bm{E}\times \bm{B}$ nonlinearity  $\alpha_{e\pm}=(1-\sqrt{2\epsilon})\langle\Phi_{\bm{k}_{\pm}}^{*}\Phi_{\bm{k}_{0}}\rangle_{s}$ via circulating electrons;  and the generalized Reynolds stress $\alpha_{i\pm}=\langle\Phi_{\bm{k}_{\pm}}^{*}(1-\langle J_{\bm{k}_{\pm}}J_{\bm{k}_{0}}J_{k_{z}}e^{iQ}\overline{e^{-iQ}}F_{0}\rangle_{v})\Phi_{\bm{k}_{0}}\rangle_{s}$ due to ions. While the former dominates in the long wavelength limit, they have comparable magnitudes for short wavelength modes.

Similarly, the pump CTEM equation are readily obtained as
\begin{eqnarray}
\label{eq:A0}
	(\partial_{t}-\gamma_{0})A_{0}=k_{z}\rho_{i}[\alpha_{+}A_{+}A^{*}_{z}-\alpha_{-}A_{-}A_{z}]/\omega_{0}\partial_{\omega}D_{0}.
\end{eqnarray}
From Eqs. (\ref{eq:Ap}), (\ref{eq:An}) and (\ref{eq:A0}), the nonlinear terms are formally independent  of trapped electrons, and  possess the conservation law of the wave energy of CTEM plasmons $E_{\bm{k}}=\omega_{\bm{k}}\partial_{\omega}D_{\bm{k}}|A_{\bm{k}}|^{2}$:
\begin{eqnarray}
\label{eq:conservation}
	(\partial_{t}-2\gamma_{0})E_{0}=(2\gamma_{+}-\partial_{t})E_{+}+(2\gamma_{-}-\partial_{t})E_{-}.
\end{eqnarray}

The  weak turbulence theory introduced so far applies for scatterings to the linearly stable domain in  both dominant \cite{chen00} and subdominant  \cite{makwana} branches. 
For a direct comparison with previous ITG results \cite{chen00,zonca04}, we now restrict consideration to the mode coupling in the same branch and analyze the modulational instability with a constant pump. Letting $\Delta_{s}=\Delta_{\pm}$, $\gamma_{s}=\gamma_{\pm}$, $D=D_{0}\simeq D_{\pm}$, and $\alpha=\alpha_{\pm}$,
Eqs.  (\ref{eq:nonlocal_Az}), (\ref{eq:Ap}) and (\ref{eq:An}) yield the desired nonlinear dispersion relation
\begin{eqnarray}
\label{eq:nonlineardispersion}
	\frac{\gamma_{m}^{2}(\gamma_{s}-\Gamma_{z})|A_{0}|^{2}\textrm{Re}(\beta_{z})}{(\Gamma_{z}+\gamma_{z})[(\Gamma_{z}-\gamma_{s})^{2}+\Delta_{s}^{2}]}=1,
\end{eqnarray}
with $\Gamma_{z}\equiv \partial_{t}$ and $\gamma_{m}^{2}\equiv 2 k^{2}_{z}\rho_{i}^{2}\alpha/(\chi_{i}\omega_{0}\partial_{\omega}D)$.
 Equation (\ref{eq:nonlineardispersion}) admits a threshold amplitude of the pump,
\begin{eqnarray}
\label{eq:threshold}
	|A_{0,c}|^{2}=\gamma_{z}\gamma_{s}(1+\Delta_{s}^{2}/\gamma_{s}^{2})/\gamma^{2}_{m}\textrm{Re}(\beta_{z}),
\end{eqnarray}
above which the ZF growth rate can be expressed, for $\Gamma_{z}>|\gamma_{s}|,\gamma_{z}$, as
\begin{eqnarray}
\label{eq:largedrive}
	2\Gamma_{z}&=& \sqrt{(\gamma_{s}+\gamma_{z})^{2}-4[\textrm{Re}(\beta_{z})\gamma_{m}^{2}|A_{0}|^{2}-\gamma_{z}\Delta^{2}_{s}/\gamma_{s}]}\nonumber\\
	& &+(\gamma_{s}-\gamma_{z}).
\end{eqnarray}
Similarities with the ITG driven zonal flow \cite{chen00} become evident.
Therefore, the CTEM turbulence and ITG turbulence are formally isomorphic in ZF excitation. The nonlinear dynamics is governed by ions and circulating electrons.
The trapped-electron contribution, on the other hand,  only enters implicitly through linear physics. For this reason, linear CTEM properties are expected to play a determinant role in the ZF excitation.
This explains the empirical observation from gyrokinetic simulations that the role of ZF appears to be connected to linear CTEM properties \cite{ernst}.

Equations (\ref{eq:threshold}) and (\ref{eq:largedrive}) also elucidate that  the nonlinear coupling coefficient $\beta_{z}$ is crucial for ZF excitation.
Noting that the spectrum of long wavelength CTEMs has a similar scale to that of ITGs \cite{adam}, it is straightforward to show that $\alpha\simeq 1-\sqrt{2\epsilon}$ and $\omega_{0}\simeq 1/(1+\tau b_{i0})$ with $b_{i0}=k_{\perp0}^{2}\rho^{2}_{i}/2\ll 1$, the coefficient $\beta_{z}$ can then be easily computed  as \cite{chen00}
\begin{eqnarray}
\label{eq:longbetaz}
	\beta_{z}\simeq-(1/\omega_{0}+\tau)k_{z}^{2}\rho_{i}^{2}/2.
\end{eqnarray}
Substituting Eq. (\ref{eq:longbetaz}) into Eqs. (\ref{eq:threshold}) and (\ref{eq:largedrive}),  one readily concludes that both the threshold $|A_{0,c}|^{2}$ and the ZF growth rate $\Gamma_{z}$ in long wavelength CTEM turbulence are  comparable to those in ITG turbulence \cite{chen00}.
That is, the CTEM-ZF interplay  occurs on a time scale similar to that of ITG turbulence, and, thus, is expected to  play important roles in saturating the long wavelength CTEM turbulence, as observed by gyrokinetic simulations \cite{ernst04,lang07,lang08,waltz08,ernst, nakata,xiao09,xiao10}.
The generic parameter dependence of ZF effects, however, is beyond the intended scope of the current study, since an adequate modelling of long wavelength CTEM spectrum is still lacking.

Conversely, the nonlinear coupling coefficients for short wavelength modes ($b_{i0}\gg 1$) can be evaluated, after some straightforward algebra,  as $\alpha\simeq 1+\tau$ and 
\begin{eqnarray}
\label{eq:shortbetaz}
	\beta_{z}\simeq-\langle\frac{\tau(1+\tau\omega_{0})k_{z}^{2}\rho_{i}^{2}}{2\sqrt{\pi}k_{\perp 0}^{3}\rho_{i}^{3}}\frac{|\Phi_{\bm{k}_{0}}|^{2}}{4\epsilon_{n}G}\frac{Z(\sqrt{-\tau\omega_{0}/4\epsilon_{n}G})}{\sqrt{-\tau\omega_{0}/4\epsilon_{n}G}}\rangle_{s},
\end{eqnarray}
which is $\mathcal{O}(\pi^{-1/2}k_{\perp 0}^{-3}\rho_{i}^{-3})$ smaller than that in Eq. (\ref{eq:longbetaz}). 
In deriving Eq. (\ref{eq:shortbetaz}), we have neglected the transit frequency in the ion propagator $\mathcal{L}_{i,\bm{k}}$,
and assumed $k_{z}^{2}\rho_{i}^{2}\ll 1$, since the short wavelength ZF has a  strong polarization shielding effect $\chi_{i}\simeq \tau $ and is more difficult to excite.
$Z$ is the  usual plasma dispersion function.
A comparison with the long wavelength CTEM case reveals that the value of $|A_{0,c}|^{2}$ is enhanced by at least $\mathcal{O}(\pi^{1/2}k_{\perp 0}^{3}\rho_{i}^{3})$ for a fixed $k_{z}^{2}\rho_{i}^{2}$. 
Physically, this phenomenon is due to the fact that, in the short wavelength limit,  the fast ion gyromotion can average out fluctuations and thus decouple ZF from CTEMs. 
Therefore, the ZF  excitation becomes much weaker in the short wavelength CTEM turbulence, consistent, again, with numerical simulation results \cite{ernst}.

Now the key issue is to determine the stability of short wavelength CTEMs.
For this purpose, we explore the eigenmode equation (\ref{eq:dispersionfunction}) in the adiabatic ion limit \cite{chen18, sm}
\begin{eqnarray}
\label{eq:eigenmode_te}
	(1+\tau)\Phi+\sqrt{2\epsilon}T\int_{\sin^{2}\frac{\theta}{2}}^{1}\frac{d\kappa^{2}\overline{\Phi}}{\sqrt{\kappa^{2}-\sin^{2}\frac{\theta}{2}}}=0,
\end{eqnarray}
where the bounce averaged potential is $\overline{\Phi}=\int_{-\theta_{b}}^{+\theta_{b}}\Phi d\vartheta/[4K(\kappa)\sqrt{\kappa^{2}-\sin^{2}(\vartheta/2)}]$,
with $K$ the complete elliptic integral of first kind and $\theta_{b}=2\sin^{-1}\kappa$ the turning point;
and $T=\{2T_{1}[\omega(1-\eta_{e}/\epsilon_{n}H)-1]+\eta_{e}(3T_{1}-1)\}/\epsilon_{n}H$,
with $T_{1}=1+\sqrt{\omega/\epsilon_{n}H}Z(\sqrt{\omega/\epsilon_{n}H})$ accounting for the precessional resonance.
Solving Eq. (\ref{eq:eigenmode_te}) numerically yields the most unstable mode $\Phi_{\bm{k}_{0}}\simeq(1+\cos\theta)/3\pi$ and an algebraic dispersion relation
\begin{eqnarray}
\label{eq:lineardispersionfunction}
	\lambda_{0}(1+\tau)+\sqrt{2\epsilon}T=0,
\end{eqnarray}
where $\lambda_{0}\simeq 1.2$, 
Assuming the linear marginal stability is achieved at a critical real frequency $\omega_{cr}$,  the threshold $\eta_{e}$ can then be derived  by setting the real and imaginary parts of Eq. (\ref{eq:lineardispersionfunction}) to zero. We find,  for modes propagating in electron diamagnetic direction ($\omega_{cr}>0$),
\begin{eqnarray}
\label{eq:etaec1}
	\eta_{e,c1}=2\epsilon_{n}/3\epsilon_{nc},
\end{eqnarray}
with $\epsilon_{nc}\equiv 2\sqrt{2\epsilon}/[3H\lambda_{0}(1+\tau)]$, and the critical real frequency $\omega_{cr1}=3(\epsilon_{n}-\epsilon_{nc})H/(2-3\epsilon_{nc}H)$.
Moreover, according to Eq. (\ref{eq:dispersionfunction}), self-consistency with the ordering $|\langle J_{0}\mathcal{L}_{i}^{-1}[(\omega+\omega_{*i}^{t})\tau F_{0}J_{0}]\rangle_{v}|\ll (1+\tau)$ requires that, in the $\tau\omega\gg \epsilon_{n}$ limit, $\omega\gg \Gamma_{0}/(1+\tau-\tau\Gamma_{0})$, with $\Gamma_{0}=I_{0}(b_{i})e^{-b_{i}}$  and $I_{0}$ a modified Bessel function. 
Including the higher-order nonadiabatic ion term, the linear growth rate near marginal stability can be obtained iteratively from Eq. (\ref{eq:dispersionfunction}) as
\begin{eqnarray}
\label{eq:imagome}
	\gamma_{0}=\frac{\textrm{Im}(T^{-1}_{1})\epsilon^{2}_{n} H^{2}}{\epsilon_{n}H-\eta_{e,c1}}[\frac{\eta_{e}-\eta_{e,c1}}{2\epsilon_{n}H}+\frac{(1+\tau\omega_{cr1})\Gamma_{0}}{2\sqrt{2\epsilon}\lambda_{0}^{-1}\omega_{cr1}}].
\end{eqnarray}
Therefore, the mode is kinetically excited via the finite toroidal precessional resonance, and the nonadiabatic ion correction is destabilizing.

The expression for $\omega_{cr1}$ implies that the positive frequency assumption will break down  for small $\epsilon_{n}<\epsilon_{nc}$. In this case, by expanding Eq. (\ref{eq:lineardispersionfunction})  around $|\omega_{cr}/(\epsilon_{n}H)|\ll 1$, one finds
\begin{eqnarray}
\label{eq:xi}
	0&=&6\epsilon_{nc}(2-4\eta_{e}+\epsilon_{n}H)\xi^{2}-3\sqrt{\pi}\epsilon_{nc}(2-3\eta_{e})\xi-2\epsilon_{n}\nonumber\\
	& &+6\epsilon_{nc}(1-\eta_{e}),
\end{eqnarray}
with $\xi\equiv\sqrt{-\omega_{cr}/(\epsilon_{n}H)}$. Thus, the CTEM considered here is a fluid-like interchange-driven instability \cite{tang},  with the following marginal stability condition
\begin{eqnarray}
\label{eq:etaec2}
	\eta_{e,c2}&=&\frac{8}{9\pi-64}\{\frac{3\pi}{4}-6+(\frac{4\epsilon_{n}}{3\epsilon_{nc}}-\epsilon_{n}H)\nonumber\\
	& &-\{[\frac{3\pi}{4}-6+(\frac{4\epsilon_{n}}{3\epsilon_{nc}}-\epsilon_{n}H)]^{2}-(\frac{9\pi}{4}-16)\nonumber\\
	& &\times[\frac{\pi}{4}-2+[\frac{4\epsilon_{n}}{3\epsilon_{nc}}(2+\epsilon_{n}H)-\epsilon_{n}H]] \}^{1/2}\}.
\end{eqnarray}
The critical real frequency is given by $\xi_{cr2}=\sqrt{\pi}(3\eta_{e}-2)/(16\eta_{e}-8-4\epsilon_{n}H)$.

From Eq. (\ref{eq:etaec1}), it is instructive to note that the system exhibits a striking feature that when $\epsilon_{n}>\epsilon_{nc}$, the fundamental threshold in Eq. (\ref{eq:etaec1}) is a critical temperature gradient rather than an $\eta_{e}$. Consequently, noting the expression of $\epsilon_{nc}$, the marginal stability condition is set by the parameters $\epsilon, \tau, r_{te}$ and $s$, consistent with numerical results \cite{ernst04, dannert,lang07,lang08,waltz08,jolliet, xiao09,xiao10}. 
On the contrary, when the density profile is sufficiently steep such that $\epsilon_{n}<\epsilon_{nc}$, the interchange-driven mode becomes unstable above a certain critical $\eta_{e,c2}$. 
Especially in the  $\epsilon_{n}\ll \epsilon_{nc}$ limit, the marginal condition (\ref{eq:etaec2}) renders $\eta_{e,c2}\simeq 1.165$ regardless of other parameters.
Note that for current parameters, the nonuniformity gradient remains weak such that one can ignore global profile variations and subdominant modes in deriving Eq. (\ref{eq:lineardispersionfunction}) \cite{sm, chen181}.

\begin{figure}[!htp]
\vspace{-0.3cm}
\setlength{\belowcaptionskip}{-0.1cm}
\centering
\includegraphics[scale=0.35]{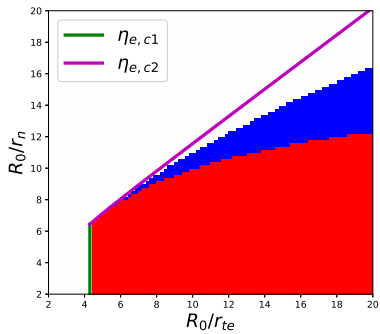}
	\caption{Stability diagram of short wavelength CTEM, for $\tau=1$, $\epsilon=0.18$ and $s=0.8$. White is stable, and blue (red) represents unstable modes propagating in the ion (electron) diamagnetic direction.}
\label{eps:stability}
\end{figure}

Confirmation that the adiabatic ion assumption captures the essential physics of short wavelength CTEMs can be obtained by solving Eq. (\ref{eq:lineardispersionfunction}) numerically and comparing the results with existing simulations. 
Figure (\ref{eps:stability}) shows the resultant stability diagram versus the density and temperature gradient scale lengths for parameters in Ref. \cite{ernst}. Three different regions can be identified: the white region is stable to short wavelength CTEMs, while the blue (red) region is unstable to modes propagating in ion (electron) diamagnetic direction.
It also validates the analytical marginal stability conditions.  
As seen in Fig. (\ref{eps:stability}), the marginal stability is indeed set by a critical $\eta_{e}$ for steep density gradient plasmas with $\epsilon_{n}<\epsilon_{nc}$, whereas it corresponds to a critical $r_{te}/R_{0}$ for $\epsilon_{n}>\epsilon_{nc}$.
Remarkably, these features are in both qualitative and quantitative agreement with simulation results (see Fig. 1  in Ref. \cite{ernst}).
Moreover, as $\eta_{e}$ moves across the threshold from the blue (red) to white region, one would expect that long wavelength modes eventually dominate the dynamics and ZFs become increasingly more important, as observed in simulations \cite{ernst}.


Under the assumption of adiabatic ions, the linear dispersion relation, Eq. (\ref{eq:lineardispersionfunction}), is independent of the toroidal mode number. In this scenario, a broad spectrum of short wavelength  CTEMs is expected,  extending  from $|k_{\theta}\rho_{i}|\gtrsim 1$ to  $|k_{\theta}\rho_{e}|\simeq \mathcal{O}(1)$, until the adiabatic  approximation of circulating electrons becomes  invalid ultimately. 
Therefore, once excited, the short wavelength CTEM turbulence characterized by weak ZF excitation will have a  broad energy-containing spectrum.

Perhaps the most significant property of short wavelength CTEMs is that they can exist with little stabilization due to zonal flows.
More specifically, noting that short wavelength CTEMs arise from the balance between the adiabatic term and the nonadiabatic trapped-electron response,
	and trapped electrons tend to bounce along magnetic field lines, the  eigenmode equation (\ref{eq:dispersionfunction}) is,  in the lowest-order adiabatic  ion limit, essentially independent of the radial modulation ($\theta_{k}$) effect. Therefore, in contrast to the ITG case, the short wavelength CTEM can be locally excited without a short radial wavelength stable domain in the dominant branch.
	Meanwhile, owing to the Hermitian nature of Eq. (\ref{eq:eigenmode_te}), the mode structures of different branches are orthogonal \cite{sm}. Thus, besides the aforementioned small $\beta_{z,\pm}$, the coefficients $\alpha_{\pm}$ vanish for the coupling between different branches, and ZFs can barely be excited in this scenario.
	Given that, as inferred from Eq. (\ref{eq:conservation}), the damping of driving mode is due to scatterings to the linearly stable domain in either dominant \cite{chen00} or subdominant \cite{makwana} branches, one can conclude that ZF excitation will not be an effective saturation channel for short wavelength CTEMs.

In summary, we have demonstrated the isomorphism between CTEM turbulence and ITG turbulence in zonal flow excitation. The nonlinear CTEM-ZF interplay is determined by ions and circulating electrons. Trapped electrons, due to their negligible finite drift orbit width and Larmor radius effects, only enter implicitly via linear physics.
The linear properties of CTEM, therefore, play a fundamental role in ZF excitation.
	While ZF scattering can be important in saturating long wavelength CTEMs, it is found that short wavelength CTEMs can exist with little stabilization due to the zonal flow excitation, and, thereby, renders the conventional drift wave-zonal flow paradigm invalid here.
Controlling parameters of the short wavelength CTEM instability are also identified analytically.
This new understanding, thus, provides a plausible explanation for the seemingly contradictory features observed in previous simulations.
	Finally, we remark that current results also carry significant implications to tokamak experiments. In particular, the short wavelength CTEM turbulence could be a candidate to explain the observed insensitivity of electron heat transport to $\bm{E}\times \bm{B}$ shearing rate as well as the decoupling of the electron thermal diffusivity  and the particle diffusion, due to the ambipolarity \cite{petty}.
	Equations (\ref{eq:etaec1}) and (\ref{eq:etaec2}), meanwhile,  can be utilized  to  identify the experimentally accessible parameters and thus control the spatiotemporal scale of CTEM turbulence.
			It may pave a way for proactively controlling  CTEM-driven electron transport and optimizing fusion plasma performance.


We acknowledge discussions with Z. Qiu, F. Zonca, E. Viezzer and M. Gracia-Munoz.
This work was supported by National Natural Science Foundation of China under Grant No. 11905097, and the European Unions Horizon 2020 research and innovation programme (No. 805162).

\section{Supplemental Material for `How Zonal Flow Affects Trapped Electron Driven Turbulence in Tokamaks'}

	In this Supplemental Material we sketch the technical details related to the eigenmode equation of short wavelength collisionless trapped-electron mode (CTEM).
	We start from the derivation of the eigenmode equation as a  special case of the CTEM equation given by \cite{chen18}.
		More specifically, by adopting the strongly ballooning representation and neglecting the nonadiabatic ion terms, 
		Eq. (2) in \cite{chen18} reduces to the following Volterra homogeneous integral equation of the second kind
\begin{eqnarray}
\label{eq:eigenmode_te0}
	(1+\tau)\Phi+\sqrt{2\epsilon}\int_{\sin^{2}\frac{\theta}{2}}^{1}\frac{T(\omega,\kappa^{2})d\kappa^{2}\overline{\Phi}}{\sqrt{\kappa^{2}-\sin^{2}\frac{\theta}{2}}}=0.
\end{eqnarray}
Here, the bounce averaged potential is $\overline{\Phi}=\int_{-\theta_{b}}^{+\theta_{b}}\Phi d\vartheta/[4K(\kappa)\sqrt{\kappa^{2}-\sin^{2}(\vartheta/2)}]$,
with $\theta_{b}=2\sin^{-1}\kappa$ the turning point of trapped electrons.
\begin{eqnarray}
\label{eq:T}
	T=\frac{2T_{1}}{\epsilon_{n}H}[\omega(1-\frac{\eta_{e}}{\epsilon_{n}H})-1]+\frac{\eta_{e}(3T_{1}-1)}{\epsilon_{n}H},
\end{eqnarray}
with $T_{1}=1+\sqrt{\omega/\epsilon_{n}H}Z(\sqrt{\omega/\epsilon_{n}H})$ accounting for the precessional resonance.
$H=4s[E(\kappa)/K(\kappa)+\kappa^{2}-1]+[2 E(\kappa)/K(\kappa)-1 ]$ describes the  pitch-angle dependence of the precessional frequency, with $K(E)$ designating the complete elliptic integral of first (second) kind.

\begin{figure}[!htp]
\vspace{-0.3cm}
\setlength{\belowcaptionskip}{-0.1cm}
\centering
\includegraphics[scale=0.35]{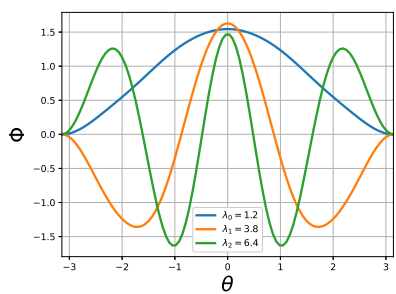}
	\caption{Plots of mode structures for different eigenvalues of Eq. (\ref{eq:eigenmode_te1}).}
\label{eps:modestructure}
\end{figure}

		Further analytic progress is possible if we approximate $H$ by its averaged value  over pitch-angle variable  $\int_{0}^{1}H d\kappa^{2}\simeq 0.83s+0.41$. 
		This approximation can be quantitatively justified in the presence of magnetic shear ($s\sim 1$) where $H$ depends very weakly on $\kappa^{2}$, and still be qualitatively relevant for weak magnetic shear plasmas \cite{adam,tang}.
		In this way, Eq. (\ref{eq:eigenmode_te0}) is rendered into 
\begin{eqnarray}
\label{eq:eigenmode_te1}
	(1+\tau)\Phi+\sqrt{2\epsilon}T\int_{\sin^{2}\frac{\theta}{2}}^{1}\frac{d\kappa^{2}\overline{\Phi}}{\sqrt{\kappa^{2}-\sin^{2}\frac{\theta}{2}}}=0.
\end{eqnarray}
		The advantage of Eq. (\ref{eq:eigenmode_te1}) is that it can be solved for an effective eigenvalue $\lambda=-\sqrt{2\epsilon}T/(1+\tau)$ without explicitly specifying parameters.
		Numerical investigations of Eq. (\ref{eq:eigenmode_te1}) are then carried out, as shown in Fig. (\ref{eps:modestructure}).
		Therefore, the resultant $\lambda$ value leads to the algebraic dispersion relation
\begin{eqnarray}
\label{eq:lineardispersionfunction_i}
	\lambda_{i}(1+\tau)+\sqrt{2\epsilon}T=0.
\end{eqnarray}
	According to the marginal stability conditions  in main text (i.e., Eqs. (18) and (21) with $\lambda_{i}$ replacing $\lambda_{0}$),  the most unstable mode corresponds to $\lambda_{0}=1.2$. Subdominant/high-order modes, such as the $\lambda_{1}$ and $\lambda_{2}$ modes  in Fig. (\ref{eps:modestructure}), are more difficult to excite  for the parameter regime considered since their fast variation along $\theta$ will be averaged over a bounce orbit.
	Nonetheless, it should be emphasized that the theoretical analyses in main text are applicable to both the dominant and subdominant modes. 
	For each $\lambda$, meanwhile,  the stability analysis in main text demonstrates that there exists only one marginal solution (with $\omega=\omega_{cr1}$ or $\omega_{cr2}$). Given that a mode becomes unstable when it moves across the real $\omega$ axis, Eq. (\ref{eq:lineardispersionfunction_i}) thus only admits one single unstable solution.
	This behavior has been confirmed numerically by using an integral method \cite{chen22}, which can locate all the eigenvalues in a given region on the complex plane.
Furthermore, another important feature which arises in the analysis of Eq. (\ref{eq:eigenmode_te1}) is the orthogonality of mode structures corresponding to different $\lambda$ values, due to the fact 
	$\langle \Phi_{j}^{*}\Phi_{i}\rangle_{s}=\lambda_{i}\int_{0}^{1} d\kappa^{2}4K\overline{\Phi}^{*}_{j}\overline{\Phi}_{i}=\lambda_{j}\int_{0}^{1}d\kappa^{2}4K \overline{\Phi}^{*}_{j}\overline{\Phi}_{i}$.

\end{document}